\documentclass[aps,prl,twocolumn,showpacs,groupedaddress]{revtex4}
\usepackage{graphicx}

\begin{document}

\draft
\title{Noise induced entanglement}
\author{J. Chen, X. X. Yi \footnote {Corresponding author},  H. S. Song,   and L. Zhou}
\affiliation{Department of physics, Dalian University of Technology, Dalian 116024, China}

\date{\today}

\begin{abstract}
We discuss the generation of entangled states of two two-level
atoms coupled  simultaneously with a dissipated atom. The
dissipation of the atom is supposed to come from its coupling to a
noise with adjustable intensity. We describe how the entanglement
between the atoms arise in such a situation, and wether a noise
except the white one could help preparation of entanglement.
Besides, we confirm that the entanglement is maximized for
intermediate values of the noise intensity, while it is a
monotonic function of the spontaneous rates. This resembles the
phenomenon of stochastic resonance and sheds more light on the
idea to exploit  noise in quantum information processing.
\end{abstract}

 \maketitle

Entanglement lies at the heart of quantum information processing
(QIP)  \cite{bennett, nielsen, plenio1}, preparation of
entanglement as a physical resource is thus a primary goal of this
field. Entangled two two-level atoms provide an ideal model for
quantum teleportation as well as a simple description for
theoretical studies in  QIP, this makes the preparation of
entangled two two-level atoms attractive and interesting.  The
main problem that must be overcome in QIP is decoherence, an
effect that  results from the coupling of the system to its
surroundings or from the inability to control precisely
experimental parameters. A consequence of decoherence is that the
entangled system may end up in a mixed state that would be no
longer useful for any quantum information processing. It is
therefore important for practical realization of quantum
information processing protocols to engineer mechanisms to
prevent, minimize, or use the impact of environmental noise.

There are a lot of proposals that have been made for preventing,
minimizing or using the environmental noise, for example, loop
control strategies, that use an ancillary system coupling to  the
quantum processor to better the performance of the
proposals\cite{wiseman,mancini}, quantum error correction
\cite{shor} uses redundant coding to protect quantum states
against noisy environments. This procedure is successful as long
as the error rate is sufficiently small. It wastes a number of
qubits and quantum gates, and then limit its implementation by
present available technology. A more economic approach consists of
exploiting the existence of so-called decoherence-free subspace
that are completely insensitive to specific types of noise
\cite{palma}. This approach tends to require fewer additional
resources, but is only applicable in specific situations. The
seminal idea that dissipation can assist the generation of
entanglement has been put forward recently \cite{plenio2,
bose,plenio3}. In a system consisting of two distinct leaky
optical cavities, it was shown that the entanglement is maximized
for intermediate values of the cavity damping rates and the
intensity of the white noise, vanishing both for small and for
large values of these parameters \cite{plenio3}. In fact, this
idea appeared first in Ref.\cite{plenio2} for two atoms inside an
optical cavity and it shows that cavity decay can assist the
preparation of maximally entangled atoms, without cavity decay,
the reduced state of the two-atom system would be in an
inseparable mixture at all times, but not in a maximally entangled
one.

In a recent paper, the idea in  Ref. \cite{plenio3}  has been put
 forward to a two-atom system \cite{yi1}, the author  used {\it  white noise} to play a
constructive role in the entanglement preparation, and shown {\it
numerically} that controllable entanglement may arise indeed in
that situation. There are two questions arise naturally, (1) if
the other kind of noise except the white one can assist
preparation of entanglement and (2) how the systems are entangled
with assistance of noise. The main goal of this paper is to answer
these questions.

Our system consists of three two-level atoms as depicted in figure
1,
\begin{figure}
\includegraphics*[width=0.95\columnwidth,
height=0.6\columnwidth]{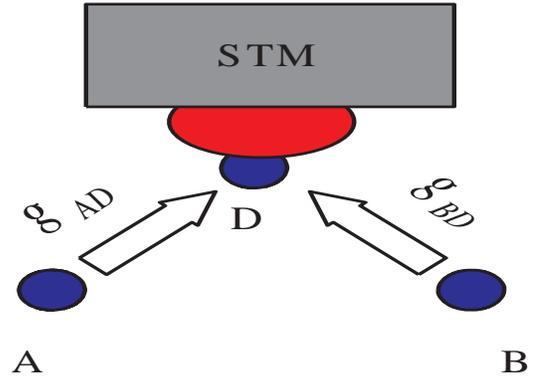} \caption{Envisaged setup
for the preparation of two entangled atoms $A$ and $B$. The atom
$D$ (Detector) is placed at the STM tip, it couples to the tip of
the STM and then is dissipated.  The three atoms are initially
prepared in their  ground state
$|g\rangle_A|g\rangle_B|g\rangle_D$. The atoms $A$ and $B$ become
entangled and the entanglement is maximized for intermediate
values of the noise intensity. } \label{fig1}
\end{figure}
we will refer to atom $A$ ,  atom $B$  and atom $D$ when the
context requires us to differentiate them, but otherwise they are
supposed to be identical. We denote the atomic ground and excited
states by $|g\rangle_i$ and $|e\rangle_i$, respectively,  and call
$2\gamma_i (i=A, B, D)$ the spontaneous emission rate from the
upper level. We will refer $\sigma_i^+=|e\rangle_i\langle g|$ to
the pauli operator for atom $i$. The atom $D$ is driven by a noise
whose intensity will be characterized by the effective particle
number $n_T$. In fact the tip connected the atom $D$ may play the
role of the noise.  The master equation governing the time
evolution of the global system is given by (setting $\hbar=1$)
\begin{equation}
\dot{\rho}=-i[H,\rho]+{\cal L} (\rho),
\end{equation}
where the Hamiltonian $H$ describes the internal energies of the
atoms as well as the inter-atom  couplings. The Liouvillean ${\cal
L }(\rho)$ describes the  decay  of atoms $A$ and $B$, as well as
the interaction of the atom $D$  with the noise. As no external
coherent driving is present, the Hamiltonian reads
\begin{equation}
H=\sum_{i=A,B,D}\frac{\omega_i}{2}\sigma^z_i+\sum_{i=A,B}
g_{iD}(\sigma_D^+\sigma_i^-+h.c.),
\end{equation}
when the noise is a white one,   the Liouvillean is given by
\cite{gardiner}
\begin{eqnarray}
{\cal L}(\rho)&=&-\gamma_D(n_T+1)(\sigma_D^+\sigma_D^-\rho+\rho
\sigma_D^+\sigma_D^- -2\sigma_D^-\rho
\sigma_D^+)\nonumber\\
&-&\gamma_D n_T(\sigma_D^-\sigma_D^+\rho+\rho
\sigma_D^-\sigma_D^+-2\sigma_D^+\rho
\sigma_D^-)\nonumber\\
&-&\sum_{i=A,B}\gamma_i(|e\rangle_i\langle e|\rho+\rho
|e\rangle_i\langle e|-2|g\rangle_i\langle e|\rho|e\rangle_i\langle
g|).\nonumber\\
\end{eqnarray}
Here $\gamma_i(i=A,B)$ describes the atom decay rate and we assume
$\gamma_A=\gamma_B=\gamma$, $n_T$ stands for the intensity of the
white noise, which  refers to its effective particle number. To
simplify the representation, now we turn to an interaction picture
with respect to $H_0=\sum_{i=A,B,D}\frac{\omega}{2}\sigma_i^z$.
After this transformation, the Liouvillean part remains unchanged,
while the Hamiltonian part is now given by
\begin{equation}
H_I=\sum_{i=A,B} g_{iD}(\sigma_D^+\sigma_i^-+h.c.) ,
\end{equation}
where we assume that the three atoms are identical, i.e., with the
same free Rabi frequency $\omega_i=\omega$. The analytical
solution to the equation (1) is extremely tedious. We will now
present numerical simulations to show how the entanglement in
system $A$ and $B$ depend on the noise intensity, time $t$ and the
decay rate of atoms $A$ and $B$.

We will choose the Wootters concurrence as the entanglement
measure \cite{wootters},
$$ c(\rho)=max\{0,\lambda_1-\lambda_2-\lambda_3-\lambda_4\},$$
where the $\lambda_i$ are the square roots of the eigenvalues of
the non-Hermitian matrix $\rho\tilde{\rho}$ with
$\tilde{\rho}=(\sigma_y\otimes
\sigma_y)\rho^*(\sigma_y\otimes\sigma_y)$ in decreasing order. The
Wootters concurrence gives an explicit expression for the
entanglement of formation, it quantifies the resources needed to
create a given entangled state. The typical behavior of the
entanglement in the system is illustrated in figure 2.
\begin{figure}
\includegraphics*[width=0.95\columnwidth,
height=0.6\columnwidth]{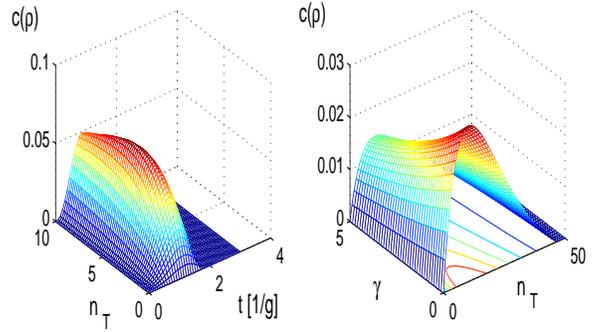} \caption{  {\it left-}
Wootters concurence of the two-atom system as a function of the
intensity of the white noise $n_T$ and time $t$. The chosen
parameters are $g_{AD}=g_{BD}=1$ and $\gamma=\gamma_D=0.2g$. {\it
right-} Wootters concurence as a function of the decay rate
$\gamma$ (namely, the decay rate for atom $A$ and $B$) and the
noise intensity $n_T$ with a specific time $t=1/g$ and
$\gamma_D=0.2g$. The other parameters are the same as the left.
Note that the entanglement arrives at its maximum for an
intermediate noise intensity in both cases.} \label{fig2}
\end{figure}
There we have plotted the amount of entanglement of the joint
state of the two atoms as a two-variable function of the intensity
of the noise $n_T$ and time $t$ (see the left panel of figure 2).
 We want to stress that our simulation
is presented for Eq.(1), i.e., the original master equation for
the three-atom system, and as we mentioned above the initial state
of the global system is $|g\rangle_A|g\rangle_B|g\rangle_D$ in our
simulation. Note that for any value of $t$ falling in the region
of entanglement $\neq 0$, the behavior of the amount of
entanglement between the two atoms is non-monotonic, it increases
to a maximum value for an optimal intensity of the noise and then
decreases towards zero for a sufficiently large intensity.
Physically, to get non-zero amount of entanglement, the excited
state of atom $D$ (act as a data bus here)  must be populated  at
least one time. For the two limiting case of either $n_T=0$ or
$n_T\rightarrow \infty$, however, the data bus remains idle for
all the times. Thus the amount of entanglement equals zero. It is
also worthwhile to study the dependence of entanglement on both
the intensity of the noise and the atom decay rates $\gamma$  of
$A$ and $B$. In the right panel of figure 2, we present those
dependence of the entanglement on $\gamma$ and $n_T$.
\begin{figure}
\includegraphics*[width=0.95\columnwidth,
height=0.6\columnwidth]{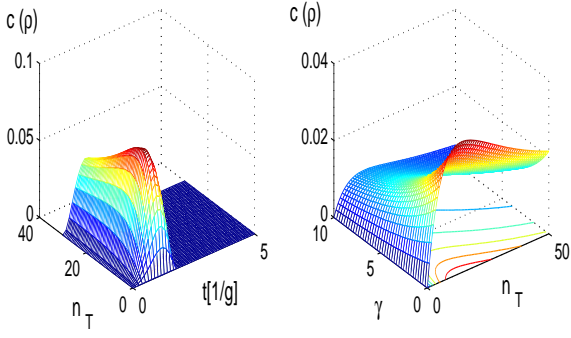} \caption{ The same as
figure 2, but for the squeezing noise.} \label{fig3}
\end{figure}
 It is interesting to note that the amount of
entanglement behave as a monotonic function of the atom decay rate
$\gamma$. This is quite different from the case presented in Ref.
\cite{plenio3}, where the cavity decay can assist themselves to
prepare entanglement shared among them.

Now we turn to study the case where the noise is a squeezed white
noise. To describe the dynamical evolution of the system under
influence of the squeezed white noise, terms
\begin{equation}
-\gamma_DM(\sigma_D^-\sigma_D^-\rho+\rho \sigma_D^-\sigma_D^-
-2\sigma_D^-\rho \sigma_D^-)+h.c.,
\end{equation}
should be added to the  Liouvillean Eq.(3). Where $M$ together
with $n_T$ characterize the noise and fulfill $M\leq
\sqrt{n_T(n_T+1)}.$ Here we will consider the ideal case in which
$M=  \sqrt{n_T(n_T+1)}$, i.e., the perfect squeezing case. The
numerical results for this situation were presented in figure 3.
By comparing figure 2 with figure 3 we find that the entanglement
are maximized for different values of $n_T$ in the two case.

To shed more light on the entanglement preparation, we now show in
a analytical way how the entanglements are generated in the
system. To make the physics clear, we introduce two new effective
atomic modes and choose the case of white noise as a
demonstration. As we will see, one of the effective modes will be
decoupled from the system. The two collective atomic modes are
given by the following definition
\begin{equation}
\sigma_C^+=\frac{g_{AD}\sigma_A^++g_{BD}\sigma_B^+}{\sqrt{g_{AD}^2+g_{BD}^2}},
\sigma_E^+=\frac{g_{BD}\sigma_A^+-g_{AD}\sigma_B^+}{\sqrt{g_{AD}^2+g_{BD}^2}},
\end{equation}
and $\sigma_i^-=(\sigma_i^+)^{\dagger}$. In terms of these new
operators, the Hamiltonian and Liouvillean part of the master
equation are given by
\begin{equation}
H_I= g_{CD}(\sigma_D^+\sigma_C^-+h.c.) ,
\end{equation}
where $g=\sqrt{g_{AD}^2+g_{BD}^2}$, and
\begin{eqnarray}
{\cal L}(\rho)&=&-\gamma_D(n_T+1)(\sigma_D^+\sigma_D^-\rho+\rho
\sigma_D^+\sigma_D^- -2\sigma_D^-\rho
\sigma_D^+)\nonumber\\
&-&\gamma_D n_T(\sigma_D^-\sigma_D^+\rho+\rho
\sigma_D^-\sigma_D^+-2\sigma_D^+\rho
\sigma_D^-)\nonumber\\
&-&\gamma\sum_{i=C,E}(|e\rangle_i\langle e|\rho+\rho
|e\rangle_i\langle e|-2|g\rangle_i\langle e|\rho|e\rangle_i\langle
g|).\nonumber\\
\end{eqnarray}
Note that the sum in the last line of Eq.(7) is taken over the two
NEW modes. The transformation between the resulting atom $A, B$
and the collective modes $C, E$ is clear. For example, both the
resulting atoms $A$ and $B$ in its ground state
$|g\rangle_A|g\rangle_B$ can be equivalently expressed in terms of
$|g\rangle_C|g\rangle_E$ and $|e\rangle_A|e\rangle_B$ likewise.
The new master equation Eq.(7) shows us that we have one mode
(mode $E$) which is completely decoupled from the Hamiltonian
dynamics and is purely damped under the Liouvillean dynamics, this
is a consequence of the transformation from the resulting atoms to
the collective modes. The mode $E$ then will not be populated in
steady state irrespective of its initial states. In other words,
if the mode $E$ is in its ground state initially, it will remain
on that forever. Therefore, we begin our investigations with both
collective modes $C$ and $E$ in the ground state
$|g\rangle_C|g\rangle_E$. As the mode $E$ will then never be
populated, we disregard that mode in the following discussions.
Apart from the above assumption, we discuss the entanglement
generation here only for the case of no initial population on the
excited state of atom $D$, this is relevant to the topics under
our consideration, i.e., study the role of the white noise in the
entanglement generation. To understand the origin of the
generation of entanglement from the white noise, let us first show
how the two modes $C$ and $D$ are entangled. To this end, we solve
perturbatively the master equation for modes $C$ and $D$ by
expanding the density matrix $\rho_{CD}(t)$ as a power of short
time $t$, it yields
\begin{equation}
\rho_{CD}(t)=\rho_{CD}(0)+\rho^{(1)}_{CD}
t+\frac{1}{2!}\rho_{CD}^{(2)} t^2+....
\end{equation}
This expansion is a good approach to the solution of the master
equation  with Hamiltonian Eq.(7) and Liouvillean Eq.(8) in the
limit $t\rightarrow 0$. Eq.(9) and Eqs (1,7,8) together give
\begin{eqnarray}
\rho_{CD}^{(1)}&=&-i[H,\rho_{CD}(0)]+{ \cal
L}(\rho_{CD}(0)),\nonumber\\
\rho_{CD}^{(2)}&=&-i[H,\rho_{CD}^{(1)}]+{\cal
L}(\rho_{CD}^{(1)}),\nonumber\\
 & \ & ...   ... \ \ \ \ ... ... \ \ \ \ ... ...
\end{eqnarray}

For a specific initial state $\rho_{CD}(0)=|gg\rangle\langle gg|$,
to the second order of time, the density matrix  $\rho_{CD}(t)$ in
the basis $\{ |gg\rangle, |ge\rangle, |eg\rangle, |ee\rangle \}$
reads
\begin{widetext}
\begin{equation}
\rho_{CD}(t)=\left( \matrix{ 1-\gamma_D n_T
t+\frac{\gamma_D^2n_T(n_T+1)+\gamma_D^2n_T^2}{2}t^2 & 0 & 0 &0 \cr
 0 &\gamma_D n_T
t-\frac{\gamma_D^2n_T(n_T+1)+\gamma_D^2n_T^2}{2}t^2
  & -\frac{ig\gamma_D n_T}{2} t^2& 0\cr
 0& \frac{ig\gamma_D n_T}{2} t^2& 0&0 \cr
 0 & 0 & 0 &0 \cr } \right).
\end{equation}
\end{widetext}
The Wootters concurence  for this state is $g\gamma_D n_T t^2$.
The physics of this result is clear, to get nonzero entanglement
in modes $C$ and $D$, either the coupling constant $g$, the decay
rate $\gamma_D$ or the noise intensity $n_T$ could not be zero,
and the entanglement increase linearly with $g$, $\gamma_D$ and
$n_T$ at the beginning of evolution. It is worthy to stress that
the entanglement shared between $C$ and $D$ do not indicate
certainly that the resulting  modes $A$ and $B$ are entangled. In
fact, it is easy to check that $A$ and $B$ are in a separable
state up to the any order of $t$. They would  not be entangled
 within the short-time approximation. Although the resulting modes
 $A$ and $B$ could not end up in an entangled state within the
 short time approximation, the short time approach provided us a manner of
 how two modes become entangled starting from their ground state
 via coupling to a noise. The entanglement share among $A$ and $B$
 arises exactly  in the same manner.

 The noise-assisted entanglement preparation is somehow reminiscent of
the well known phenomenon of stochastic resonance
\cite{gammaitoni,huelga,buchleitner}, where the response of a
system to a periodic force can be enhanced in the presence of an
intermediate amount of noise. A related effect that cavity decay
can assist the generation of squeezing has been found recently
\cite {nha}, there they shown that the squeezing effect is
enhanced as the damping rate of the cavity is increased to some
extent, and the pumping field amplitude is required to be
inversely proportional to the damping rate for the optimal
squeezing.

As an example, we now describe a setup for entangled atom pair
creation with two $^{31}P$ ions deposited at the (111) surface of
$^{28}Si$ substrate. As the author did in Ref. \cite{long}, we
choose $^{31}P$ that has nuclear spin $I=\frac 1 2 $ act as the
two-level system. A $^{13}C$ was put on the tip of STM, the
$^{31}P$ ions in the substrate was deposited with a large distance
such that no direct interaction between them. With the STM tip
approaches the two $^{31}P$ ions, the entanglement starts to be
generated and the two $^{31}P$ ions might end up in an entangled
state.

To sum up, we have described an experimental situation where
entanglement between two atomic systems can be prepared with
assistance of a  noise, the noise might  be  a white one or a
squeezed white one. The entanglement measured by the Wootters
concurence is maximized for intermediate values of the intensity
of the  noise, while it is a monotonic function of the atomic
spontaneous emission rate. Recall that the atomic decay itself can
not induce entanglement among the atoms, even if at finite
temperatures, we conclude that the coupling between the data bus
and the white noise is the origin of the generation of the
entanglement. The phenomenon of white noise-assisted entanglement
generation is not a rare phenomenon, it resembles the phenomenon
of stochastic resonance. However, this discovery \cite{plenio3} is
really valuable because it sheds new light on the constructive
role that noise may play in quantum information processing. In
contrast with the results in Ref.\cite{plenio3}, the proposal
presented here is for the entanglement generation between two
two-level atoms. For such a two-qubit system, any amount of
entanglement, even if very small, is distillable \cite{yi2}, and
therefore the entangled atoms are useful for quantum information
processing. \ \ \\
\ \ \\
This work is supported
by EYTP of M.O.E, and NSF of China.\\


\begin{references}
\bibitem{bennett} C. H. Bennett and D. P. DiVincenzo, Nature {\bf
404}, 247(2000).

\bibitem{nielsen}  M. A. Nielsen and I. L. Chuang, Quantum computation and
quantum information (Cambridge University press, Cambridge, 2000).
\bibitem{plenio1} M. B. Plenio and V. Vedral, Contemp. Phys. {\bf
39},431(1998).

\bibitem{wiseman} H. M. Wiseman and G. J. Milburn, Phys. Rev.
Lett. {\bf 70},548(1993); D. Vital, {\it et al.,} Phys. Rev.
Lett.{\bf 79}, 2442(1997); L. Viola,{\it et al.,} Phys. Rev. Lett.
{\bf 82}, 2417(1999).
\bibitem{mancini} S. Mancini, {\it et al.,} EuroPhys. Lett. {\bf 60}, 498(2002).
\bibitem{shor} P. W. Shor, Phys. Rev. A {\bf 52}, 2493(1995); A.
R. Calderbank and P. W. Shor, Phys. Rev. A {\bf 54}, 1098(1996);
A. M. Steane, Proc. Roy. Soc. A {\bf 452}, 2551(1996).
\bibitem{palma} G. M. Palma, {\it et al.,} Proc. Roy. Soc. A {\bf
452}, 567(1996); M. B. Plenio, {\it et al.,} Phys. Rev. A {\bf
55}, 67(1997); D. A. Lidar, {\it et al.,} Phys. Rev. Lett. {\bf
81}, 2594(1998); A. Beige, {\it et al.,} Phys. Rev. Lett. {\bf
85}, 1762(2000); A. Beige, {\it et al.,} New J. Phys. {\bf 2},
22(2000).
\bibitem{plenio2} M. B. Plenio, {\it et al.,} Phys. Rev. A {\bf
59}, 2468(1999); A. Beige, {\it et al.,} J. Mod. Opt. {\bf 47},
2583(2000); P. Horodecki, Phys. Rev. A {\bf 63},022108(2001).
\bibitem{bose} S. Bose, {\it et al.,}Phys. Rev. Lett. {\bf 83},
5158(1999); D. Braun, Phys. Rev. Lett. {\bf 89}, 277901(2002).
\bibitem{plenio3} M. B. Plenio, {\it et al.,} Phys. Rev. Lett.
{\bf 88}, 197901(2002).
\bibitem{yi1} X. X. Yi, {\it et al.,} quant-ph/0306091; to appear in PRA.
\bibitem{gardiner} C. W. Gardiner and P. Zoller, Quantum Noise
(Springer 2000).

\bibitem{wootters} W. K. Wootters, Phys. Rev. Lett. {\bf 80},
2245(1998).
\bibitem{gammaitoni} L. Gammaitoni, {\it et al.,} Rev. Mod. Phys.
{\bf 70}, 223(1998).
\bibitem{huelga} S. F. Huelga and M. B. Plenio, Phys. Rev. A {\bf
62}, 052111(2000).
\bibitem{buchleitner} A. Buchleitner, and R. N. Mantegna, Phys.
Rev. Lett.{\bf 80}, 3932(1998).
\bibitem{nha} H. Nha, {\it et al.,} Phys. Rev. A {\bf 67}, 023801(2003).
\bibitem{long} G. L. Long {\it et al.,} quant-ph/0307054.
\bibitem{yi2} M. Horodecki, {\it et al.,} Phys. Rev. A {\bf 59}, 4206
(1999); X. X. Yi,{\it et al.,}  Phys. Rev. A {\bf 62},
062312(2000).
\end{references}
\end{document}